\documentclass{kluwer} \usepackage{epsfig}   

\newdisplay{guess}{Conjecture}

\begin{document}                                                                                   
\begin{article}
\begin{opening}         
\title{The thin layer of Warm Ionized Gas: towards a 3-D reconstruction of the 
spatial distribution of HII regions} 
\author{Roberta \surname{Paladini}$^{1}$, Rod \surname{Davies}$^{2}$, Gianfranco \surname{DeZotti}$^{3}$}  
\runningauthor{Roberta Paladini, Rod Davies, Gianfranco DeZotti}
\runningtitle{The thin layer of Warm Ionized Gas: towards the 3-D 
reconstruction of the spatial distribution
              of Galactic HII regions}
\institute{$^{1}$SISSA, International School for Advanced Studies, via Beirut 2-4, 
I-34014 Trieste, Italy\\
$^{2}$University of Manchester, Jodrell Bank Observatory, Macclesfield - Cheshire SK11 9DL, 
UK\\
$^{3}$INAF-Oss. Astro. Padova, Vicolo dell'Osservatorio 5, I-35122 Padova, Italy}

\begin{abstract}
HII regions are known to contribute to the so-called {\em thin layer}
of the diffuse Warm Ionized Gas. In order to constrain this
contribution, we reconstruct the 3-D distribution of the sources.
A detailed spatial analysis of the largest up-to-date sample of HII regions is presented. 
\end{abstract}
\keywords{WIM, HII regions}
\end{opening}           

\section{Introduction}  

The thin disk component of the Galactic free electron distribution is largely contributed by 
localized HII regions. In order to constrain this contribution, it is important 
to reconstruct the spatial distribution of known sources.\\
In a recent paper (Paladini et al., 2002, hereafter Paper I), we describe   
the construction of an extensive radio catalog (1442 sources) of Galactic HII regions
by the combination of 24 published lists and catalogs of these objects.
The final compilation consists of a Master Catalog
(containing original data and corresponding errors) and
a Synthetic Catalog at 2.7 GHz (which summarizes the basic information -
flux density, angular diameter and $V_{LSR}$ - for each source of the Master Catalog).
The Synthetic Catalog has provided the source of information for the spatial
analysis here presented. 

\section{3-D Distribution of Galactic HII regions}

In order to  assess the level of completeness of our sample we study the
derived log N - log S distribution (Fig.~1).

\begin{figure}
\begin{center}
\epsfig{figure=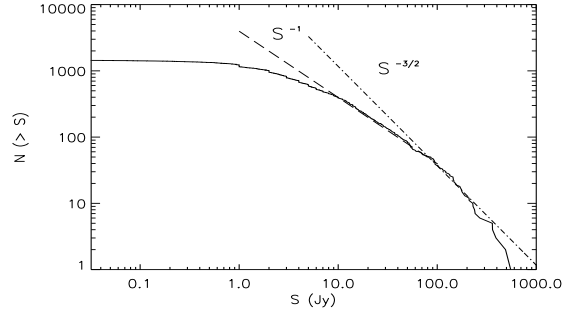,
height=4.5cm, width=7.5cm,angle=0}
\caption{Cumulative counts N($> S$) for the 2.7 GHz Synthetic Catalog.}
\end{center}
\end{figure}

\noindent
This can be well fitted by a two-component power-law such as:

\begin{equation}
 N(>S) \propto \left \{ \begin{array}{ll}
            S^{-\frac{3}{2}} & S_{max} > S>  \hspace{0.1truecm} \sim  \hspace{0.1truecm} 70 
\hspace{0.1truecm} Jy \nonumber \\
            S^{-1} & \sim  \hspace{0.1truecm} 70  \hspace{0.1truecm} Jy > S >  \hspace{0.1truecm} \sim 1 
\hspace{0.1truecm} Jy
              \end{array} \right.
\end{equation}

\noindent
For $S_{max} > S> \sim$ 70 Jy  we are sampling well into the
local spiral arm (spherical distribution approximation) while for $S_{max} > S> 
\sim$ 70 Jy we sample into more distant regions into the Galactic disk (disk 
distribution approximation). Below $\sim 7$ Jy, the Synthetic Catalog 
starts missing sources which are in regions of high confusion level.\\
Velocities are given for $\sim$ 800 of the 1442 HII regions in the Synthetic
Catalog, corresponding to $60\%$ of the total. Details about the kinematic data are
given in Paper I.
These data have been combined with the linear  rotation model by Fich, Blitz and Stark 1989:

\begin{equation}
\Theta = (221.64 - 0.44 R) \hspace*{0.1truecm} km \hspace*{0.1truecm} s^{-1}
\end{equation}

\noindent
to compute galactocentric (R) and solar distances (D). Taking into account the
typical measurement error on radial velocities (a few km s$^{-1}$), we 
have removed from our sample all the sources with an observed velocity $< |10|$ km s$^{-1}$.
These sources are characterized by a large distance-uncertainty. The recovered 
radial distribution is shown in Fig.~2.

\begin{figure}
\begin{center}
\epsfig{figure=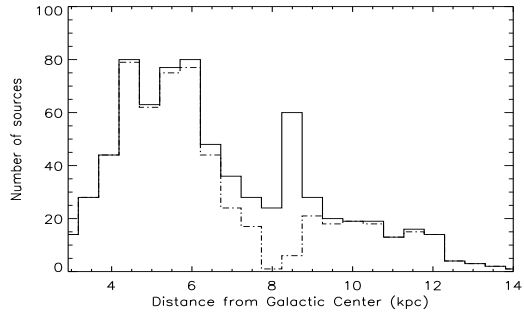,
height=4.5cm, width=7.5cm, angle=0}
\caption
{Radial distribution of the Synthetic Catalog HII regions (solid line).
Overlaid (dashed-dotted line) is the
distribution obtained by considering
only sources with velocities $>|10|$ Km $s^{-1}$.}
\end{center}
\end{figure}

\noindent
In computing solar distances, one has to consider the well known problem
of distance degeneracy for sources lying inside the solar circle. Auxiliary data can be useful 
in order to solve this ambiguity. In particular, we have used absorption lines data (HI, H$_{2}$CO) 
and catalogued optical counterparts. HI data have been mainly taken from
Kuchar $\&$ Bania 1994 and Caswell et al. 1975. Additional data
are from Kerr $\&$ Knapp 1970 and Goss $\&$ Radhakrishnan 1969.
H$_{2}$CO data are from Wilson 1980. The optical catalogs which
have been used are the Mars\'alkov\'a 1974 Master Catalog, the Blitz et al. 1982 Catalog
and the Brand $\&$ Blitz 1993 Catalog. Complementary data on individual sources are from
Miller 1968 and Shaver et al. 1981. Through this method, we are able to assign a 
solar distance to 177 sources. These sources have to be added to 143 for which a
unique distance can be computed from kinematic data. For the remaining 288 sources, we
have worked out a method based on a distance indicator independent from kinematic data. 
This distance indicator (whose robustness is currently under analysis) has been found from a 
luminosity vs. physical diameter
correlation obtained by exploiting the 2.7 and 5 GHz flux density and angular diameter data
of the Master Catalog. Therefore, according to this
correlation, we can compute the solar distance as:

\begin{equation}
D = \left(\frac{10^{a}*(\theta/206265 \times 10^{3})^{b}}{4 \pi S_\nu} \right)^{\frac{1}{2-b}}
\end{equation}

\noindent
where the recovered values of the parameters $a$ and $b$ are, respectively, $\sim$ 30.6 
and 1.1. The correlation-method turns out to be appliable
to 256 HII regions ($\sim$ 89$\%$ of the total number of sources with a distance ambiguity)
for which 
at least one observed value of flux density and angular diameter at 2.7 or
5 GHz is available. With these computed solar distances, we are also able to determine the thickness of 
the Galactic HII regions layer. A preliminary analysis has retrieved a value in the range $\sim$ 32-51 pc, 
depending on the inclusion, in the calculation, of only sources with a safely defined solar distance
(former lower value) or also of sources with a correlation-assigned distance (latter one).

\acknowledgements
R. Paladini acknowledges financial support from ESA for
the participation in the workshop.

\begin{figure}
\begin{center}
\caption{2-D distribution of HII regions from the Synthetic Catalog at 2.7 GHz.
The plot combines: 143 unambiguous sources (diamond) ; 177 sources having either optical or 
absorption auxiliary information (diamond) ; 256 {\it degenerate}
sources for which near (triangle) or far (cross) solar distances have been
assigned through the correlation method.
Overlaid on the 2-D distribution is the spiral arms model
by Taylor $\&$ Cordes 1993. 
}
\vspace*{1truecm}
 \epsfig{figure=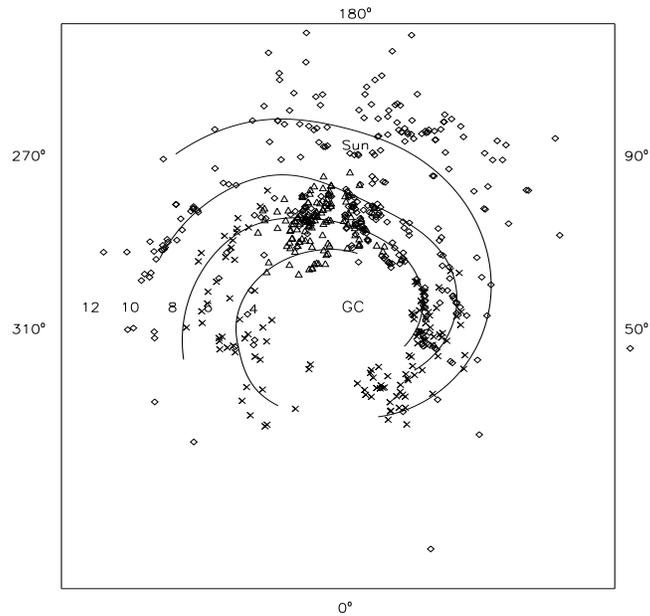,
height=8cm,width=9cm,angle=0}
\end{center}
\end{figure}

\end{article}
\end{document}